\documentclass[epj]{svjour}

\usepackage{dsfont}
\usepackage{graphicx}
\usepackage{epstopdf}  

\begin{document}
\title{The nucleon as a test case to calculate vector-isovector form factors at low energies}
\author{Stefan Leupold
}                     
\authorrunning{S.\ Leupold}
\titlerunning{Nucleon form factors}
\institute{Institutionen f\"or fysik och astronomi, Uppsala Universitet, Box
516, S-75120 Uppsala, Sweden}
%
\date{06.09.2017}
%

\abstract{
Extending a recent suggestion for hyperon form factors to the nucleon case, dispersion theory is used to relate the low-energy 
vector-isovector form factors of the nucleon to the pion vector form factor. 
The additionally required input, i.e.\ the pion-nucleon scattering amplitudes 
are determined from relativistic next-to-leading-order (NLO) baryon chiral perturbation theory 
including the nucleons and optionally the Delta baryons. Two methods how to include pion rescattering are compared: (a) solving 
the Muskhelishvili-Omn\`es (MO) equation and (b) using an N/D approach. It turns out that the results differ strongly from 
each other. Furthermore the results are compared to a fully dispersive 
calculation of the (subthreshold) pion-nucleon amplitudes based on Roy-Steiner (RS) equations. 
In full agreement with the findings from the hyperon sector it turns out that the inclusion of Delta baryons is 
not an option but a necessity to obtain reasonable results. The magnetic isovector form factor depends strongly on a low-energy 
constant of the NLO Lagrangian. If it is adjusted such that the corresponding magnetic radius is reproduced, then the 
results for the 
corresponding pion-nucleon scattering amplitude (based on the MO equation) agree very well with the RS results. 
Also in the electric sector the Delta degrees of freedom are needed to obtain the correct order of magnitude for the isovector 
charge and the corresponding electric radius. Yet quantitative agreement is not achieved. 
If the subtraction constant that appears in the 
solution of the MO equation is not taken from nucleon+Delta chiral perturbation theory but adjusted such that the electric 
radius is reproduced, then one obtains also in this sector a pion-nucleon scattering amplitude that agrees well with 
the RS results. 
} 
\maketitle

\section{Introduction and Summary}
\label{sec:intro}

The remaining big challenge within the standard model of particle physics is to understand quantitatively how the quarks 
and gluons form nucleons and other hadrons. Especially when light quarks are involved this means that we have to understand 
the formation of composite objects from relativistic quantum states. 
In general, form factors parametrize the deviation from pointlike behavior \cite{pesschr}. Thus they encode by 
definition the information 
about the intrinsic structure of an object. On the other hand, when quantum objects are probed in relativistic reactions, 
quantum fluctuations influence the measurement. In this sense there are no truly pointlike objects in the realm of relativistic
quantum physics. These quantum fluctuations are nothing but the cross-channel equivalent of particle production. On a technical
level the analyticity of reaction amplitudes enforces the presence of quantum fluctuations whenever the optical theorem 
incorporates the corresponding particle production. Dispersion theory is the natural framework to establish these 
interrelations \cite{zbMATH03081975,Omnes:1958hv,bjorkendrell}. 

The lower the energy/momentum that one uses to probe the object of interest, the less resolution one can achieve. Consequently 
electromagnetic form factors coincide in the low-energy limit with the properties that one attributes already to a 
pointlike object, the electric charge and magnetic moment. Proceeding to somewhat higher energies one can measure the onset 
of an energy dependence of a form factor. In a dispersive representation this is related to the 
lightest particles that 
couple to the object of interest and to electromagnetism. The impact of heavier states is suppressed by the (square of the)
ratio between the resolution energy and the heavy mass of these states. The lightest hadronic state that couples to 
electromagnetism is the two-pion state \cite{pdg}. 
The dispersive framework \cite{Frazer:1960zzb,Hohler:1976ax,Mergell:1995bf,Hoferichter:2016duk} 
that utilizes these interrelations is at the heart of the present work. 

Recently it has been proposed in \cite{Granados:2017cib} to determine the low-energy electromagnetic form 
factors for the transition of the Sigma to 
the Lambda hyperon by a combination of dispersion theory and relativistic octet+decuplet chiral perturbation theory ($\chi$PT) at 
next-to-leading order (NLO). A similar framework --- with subtle differences that will be addressed below --- has been used in 
\cite{Alarcon:2017asr} based on relativistic octet+decuplet $\chi$PT at leading order (LO). 
In the latter work, peripheral transverse densities for the whole ground-state multiplet have been determined.
Recently an extension of the framework of \cite{Alarcon:2017asr} to the scalar form factor of the nucleon using NLO $\chi$PT 
has been presented in \cite{Alarcon:2017ivh}. 

In \cite{Granados:2017cib} the use of $\chi$PT has been motivated by the fact that there exist no direct 
pion-hyperon scattering data. Clearly this situation is different for the nucleon case. There a dispersive analysis based on 
Roy-Steiner equations exists for the pion-nucleon scattering amplitudes \cite{Hoferichter:2015hva}. It can be used to pin down
the subthreshold t-channel p-wave pion-nucleon amplitudes. In turn these subthreshold amplitudes provide the necessary input for 
the vector-isovector form factors of the nucleon \cite{Hoferichter:2016duk}. 

Thus one might use the nucleon case 
as a cross-check of the formalism proposed for the hyperons in \cite{Granados:2017cib}. This was the primary motivation to start 
the present work. In addition the nucleon case is of course interesting in its 
own right \cite{Punjabi:2015bba,Pohl:2010zza,Carlson:2015jba,Hoferichter:2016duk}. Finally it is worth to compare the 
two approaches \cite{Granados:2017cib} and \cite{Alarcon:2017asr}. Though both respect Watson's theorem of the universality of 
final-state interactions \cite{Watson:1954uc} pion rescattering is treated very differently. 
A Muskhelishvili-Omn\`es (MO) problem \cite{zbMATH03081975,Omnes:1958hv} is solved in \cite{Granados:2017cib} 
while a variant of the N/D method \cite{PhysRev.119.467} 
is utilized in \cite{Alarcon:2017asr}. It will turn out that the results are very different with the MO version agreeing very well
with the fully dispersive setup of the Roy-Steiner analysis. 

The analysis presented in the following supports the ideas raised in \cite{Granados:2017cib} for hyperons: The explicit inclusion
of decuplet degrees of freedom (for the nucleon case the Delta baryon) is mandatory. 
Undetermined parameters can be fitted to data. Consequently the required input is
\begin{itemize}
\item once subtracted dispersion relations for the magnetic and electric isovector form factors, the respective subtraction 
  constant is fixed by the corresponding magnetic moment or charge;
\item the by now very well known pion vector form factor \cite{Hanhart:2012wi,Schneider:2012ez,Hoferichter:2014vra} based on 
  the pion p-wave phase shift \cite{GarciaMartin:2011cn,Colangelo:2001df};
\item the exchange diagrams of octet and decuplet baryons with coupling constants adjusted to pertinent data on 
  pion-baryon interactions;
\item for each sector (electric/magnetic) a constant --- pion-baryon contact interaction --- that enters the solution of the 
  MO problem, this constant can be fitted to the corresponding radius.
\end{itemize}
Thus, with value and slope at the photon point as input, the shape of a form factor, e.g.\ its curvature can be predicted. 
Alternatively one might use the obtained dispersive representation with free parameters to fit 
form factor data. This is similar in spirit 
to \cite{Abouzaid:2009ry}.\footnote{I thank Emilie Passemar for suggesting this idea.} 
In that way a parametrization superior to polynomial fits might be obtained that 
correctly accounts for pion rescattering and for close-by left-hand cuts. Of course, a description of data on the scattering of 
electrons on protons or neutrons requires the additional treatment of the isoscalar part of the electromagnetic form factors.
This is beyond the scope of the present work where I solely focus on the isovector part.

\section{Dispersive framework}
\label{sec:disp}

Essentially I follow the formalism described in \cite{Granados:2017cib}. 
To apply dispersion theory I formally study the (isovector part of the) reaction $N \, \bar N \to \gamma^*$ and saturate the 
intermediate states by a pion pair. It can be expected that the saturation of the inelasticity by a pion pair provides a good 
approximation for the form factors at low energies. 

The form factors can be defined (in the isospin limit) by 
\begin{eqnarray}
  && \langle 0 \vert j^\mu \vert (p \bar p - n \bar n)/2 \rangle \nonumber \\ 
  && = e \, \bar v_N \, \left( \gamma^\mu \, F_1(q^2) - \frac{i \sigma^{\mu\nu} \, q_\nu}{2 m_N} \, F_2(q^2) 
  \right) \, u_N
  \label{eq:defFF}
\end{eqnarray}
with
\begin{eqnarray}
  G_E(q^2) & := & F_1(q^2) + \frac{q^2}{4 m_N^2} \, F_2(q^2) \,, 
  \nonumber \\ 
  G_M(q^2) & := & F_1(q^2) + F_2(q^2)  \,.
  \label{eq:defFFEM}
\end{eqnarray}
$q^2$ denotes the square of the invariant mass of the virtual photon. With the conventions of (\ref{eq:defFF}) the photon 
momentum $q$ is given by the sum of the momenta of the two baryons.
In the following $G_{E/M}$ is called electric/magnetic form factor, i.e.\ I will not
always stress explicitly that these are only the isovector parts of the commonly known electromagnetic form factors.
It is worth to mention that $G_M$ is the helicity flip and $G_E$ the helicity non-flip amplitude concerning the 
baryon spins in the 
reaction $N \, \bar N \to \gamma^* \to e^+ e^-$; see also \cite{Korner:1976hv}.

I will mainly use the subtracted dispersion relations
\begin{eqnarray}
  &&  G_{M/E}(q^2) =  G_{M/E}(0)  
  \nonumber \\ && {}
  +  \frac{q^2}{12\pi} \, \int\limits_{4 m_\pi^2}^\infty \frac{ds}{\pi} \, 
  \frac{T_{M/E}(s) \, p_{\rm c.m.}^3(s) \, F^{V*}_\pi(s)}{s^{3/2} \, (s-q^2-i \epsilon)}  \,. \phantom{m}
  \label{eq:dispbasic}  
\end{eqnarray}
The subtraction constants that appear in (\ref{eq:dispbasic}) can be adjusted to match the form factors at the 
photon point, $G_E(0)=\frac12$, $G_M(0)=\frac12(1+\kappa_p-\kappa_n)$ where $\kappa_{p/n}$ denotes the magnetic moment of the 
proton/neutron. 

In line with the names for the form factors I will denote 
the corresponding pion-nucleon amplitudes $T_E$ and $T_M$ by electric and magnetic scattering amplitude, 
respectively. These quantities are reduced amplitudes for the formal reaction 
$N\,\bar N$ $\to$ $\pi^+\,\pi^-$ projected on $I=1$, $J=1$. For details I refer again to \cite{Granados:2017cib}.
Yet, to make comparisons to other works easier I will relate $T_{M/E}$ to the amplitudes 
used in \cite{PhysRev.117.1603,Becher:2001hv,Hoferichter:2016duk}. This matching is described in appendix \ref{sec:app0}.

In (\ref{eq:dispbasic}) $p_{\rm c.m.}$ denotes the pion momentum in the center-of-mass frame of the two-pion system and 
$F^{V}_\pi$ the pion vector form factor defined by
\begin{equation}
  \label{eq:pionFFdef}
  \langle 0 \vert j^\mu \vert \pi^+(p_+) \, \pi^-(p_-) \rangle = e \, (p_+^\mu -p_-^\mu) \, F^V_\pi((p_++p_-)^2)  \,. \phantom{m}
\end{equation}

Besides the once subtracted dispersion relation (\ref{eq:dispbasic}) I will also examine an unsubtracted version,
\begin{eqnarray}
  G_{M/E}(q^2) = 
  \frac{1}{12\pi} \, \int\limits_{4 m_\pi^2}^\infty \frac{ds}{\pi} \, 
  \frac{T_{M/E}(s) \, p_{\rm c.m.}^3(s) \, F^{V*}_\pi(s)}{s^{1/2} \, (s-q^2-i \epsilon)}  
  \label{eq:dispbasicunsubtr}  
\end{eqnarray}
and explore to which extent the pion loop plus pion rescattering saturates the isovector magnetic moment,
\begin{eqnarray}
  \frac12(1+\kappa_p-\kappa_n) \stackrel{?}{=} 
  \frac{1}{12\pi} \, \int\limits_{4 m_\pi^2}^\infty \frac{ds}{\pi} \, 
  \frac{T_{M}(s) \, p_{\rm c.m.}^3(s) \, F^{V*}_\pi(s)}{s^{3/2} }  \,, \phantom{n}
  \label{eq:dispbasicunsubtrkappa}  
\end{eqnarray}
and to which extent the dispersively calculated isovector charge is reproduced,
\begin{eqnarray}
  \frac12 \stackrel{?}{=} 
  \frac{1}{12\pi} \, \int\limits_{4 m_\pi^2}^\infty \frac{ds}{\pi} \, 
  \frac{T_{E}(s) \, p_{\rm c.m.}^3(s) \, F^{V*}_\pi(s)}{s^{3/2} }  \,.
  \label{eq:dispbasicunsubtrnull}  
\end{eqnarray}
Concerning the quality of subtracted vs.\ unsubtracted dispersion relations I refer to the detailed discussion 
in \cite{Granados:2017cib} and references therein. The synopsis is that a subtracted dispersion relation is more 
reliable than an unsubtracted one if one keeps from all possible intermediate states only the ones that remain relevant at 
low energies, i.e.\ the two-pion states; see also the corresponding discussion in \cite{Hoferichter:2016duk}. 

In line with \cite{Hoferichter:2016duk} I also introduce electric and magnetic radii:
\begin{eqnarray}
  \label{eq:defradiusEME}
  \langle r^2_{M/E} \rangle := 6 \left. \frac{d G_{M/E}(q^2)}{dq^2} \right\vert_{q^2 = 0} \,.
\end{eqnarray}
These isovector radii are related to the standard electromagnetic radii $\langle r^2_{M/E} \rangle_{p/n}$ for proton and 
neutron via
\begin{eqnarray}
  \label{eq:radiipn}
  \langle r^2_{M} \rangle & = & \frac12 \, 
  \left( (1+\kappa_p) \, \langle r^2_{M} \rangle_p - \kappa_n \, \langle r^2_{M} \rangle_n \right)  \,, \\
    \langle r^2_{E} \rangle & = & \frac12 \, 
  \left( \langle r^2_{E} \rangle_p - \langle r^2_{E} \rangle_n \right)  \,.
\end{eqnarray}
As a consequence of (\ref{eq:dispbasic}) the dispersive representation of the radii reads
\begin{eqnarray}
  \langle r^2_{M/E} \rangle = 
  \frac{1}{2\pi} \, \int\limits_{4 m_\pi^2}^\infty \frac{ds}{\pi} \, 
  \frac{T_{M/E}(s) \, p_{\rm c.m.}^3(s) \, F^{V*}_\pi(s)}{s^{5/2}}  \,.
  \label{eq:dispbasicradii}  
\end{eqnarray}

To satisfy Watson's theorem \cite{Watson:1954uc} the amplitudes $T_{M/E}$ must contain the rescattering of pions. I will discuss 
two unitarization methods how to account for this rescattering. The solution of the 
Muskhelishvili-Omn\`es problem \cite{zbMATH03081975,Omnes:1958hv} provides the basis of the first approach. 
I will denote the results by $T^{\rm MO}$ and 
suppress the labels $M/E$ until they become 
relevant again. The resulting form factors are denoted by $G^{\rm MO}$. 
This MO approach has also been used in \cite{Granados:2017cib}. 

The second framework is a variant of the N/D method \cite{PhysRev.119.467}. 
It has been used in \cite{Alarcon:2017asr} and is based on a further rewriting of the imaginary part of the form factors, 
i.e.\ of the numerators in (\ref{eq:dispbasic}), (\ref{eq:dispbasicunsubtr}).
I will denote the solution by $G^{\rm N/D}$. 

For the MO framework the amplitude $T^{\rm MO}$ is decomposed into one part that contains all the left-hand cuts and another 
that contains the right-hand cuts. 
The former is denoted by $K$. For the problem at hand where there are no overlapping cuts 
one finds:
\begin{eqnarray}
  \label{eq:startMO}
  {\rm Im}(T^{\rm MO}-K) = T^{\rm MO} \, e^{-i\delta} \, \sin\delta
\end{eqnarray}
with the pion p-wave phase shift $\delta$. Equation (\ref{eq:startMO}) is solved by the ansatz
\begin{eqnarray}
  \label{eq:ansatzMO}
  T^{\rm MO}-K = \Omega \, H
\end{eqnarray}
with an auxiliary function $H$. The Omn\`es function
\begin{eqnarray}
  \Omega(s) = \exp\left\{ s \, \int\limits_{4m_\pi^2}^\infty \frac{ds'}{\pi} \, \frac{\delta(s')}{s' \, (s'-s-i \epsilon)} \right\}
  \label{eq:omnesele}  
\end{eqnarray}
solves the homogeneous version of (\ref{eq:startMO}), i.e.
\begin{eqnarray}
  \label{eq:startMOhom}
  {\rm Im}\Omega = \Omega \, e^{-i\delta} \, \sin\delta  \,.
\end{eqnarray}
Note that by construction both $H$ and $\Omega$ have no left-hand cuts, but only the right-hand cut from the two-pion states. 
After some rewriting one obtains:
\begin{eqnarray}
  \label{eq:eqH}
  0 = {\rm Im}(T^{\rm MO}-K) - T^{\rm MO} \, e^{-i\delta} \, \sin\delta  \nonumber \\
  = e^{-i\delta} \, \left( \vert\Omega\vert \, {\rm Im}H - K \, \sin\delta \right) 
\end{eqnarray}
and therefore
\begin{eqnarray}
  \label{eq:eqH2}
  {\rm Im}H = \frac{K \, \sin\delta}{\vert\Omega\vert} \,.
\end{eqnarray}
Following still \cite{Granados:2017cib} I determine $H$ from a subtracted dispersion relation by recalling that $H$ 
has only a right-hand cut (caused by the two-pion states):
\begin{eqnarray}
  H(s) & = & P_{n-1}(s) \nonumber \\ && {}+ s^n \, \int\limits_{4m_\pi^2}^\infty \, \frac{ds'}{\pi} \, 
    \frac{\sin\delta(s') \, K(s')}{\vert\Omega(s')\vert \, (s'-s-i \epsilon) \, {s'}^n}  \,. \phantom{mm}
  \label{eq:eqH3}
\end{eqnarray}
Here $P_m$ is a polynomial of degree $m$. In practice I use $n=1$. 
Finally this yields 
\begin{eqnarray}
  T^{\rm MO}(s) & = & K(s) + \Omega(s) \, P_0 \nonumber \\ && {}+ \Omega(s) \, s \, 
    \int\limits_{4m_\pi^2}^\infty \, \frac{ds'}{\pi} \, 
    \frac{\sin\delta(s') \, K(s')}{\vert\Omega(s')\vert \, (s'-s-i \epsilon) \, {s'}} \,. \phantom{mm}
  \label{eq:tmandel}
\end{eqnarray}
Though the whole setup deals with and aims at low-energy quantities, it is nonetheless interesting to note 
that the use of a constant $P_0$ instead of a higher-order polynomial $P_n(s)$, $n\ge 1$, has the following appealing 
feature. Assuming $\Omega(s) \sim 1/s$ (see, e.g., \cite{Kang:2013jaa}) and $K(s)$ dropping for large $s$, 
then also $T^{\rm MO}$ drops. This leads to a convergent
integral in (\ref{eq:dispbasicunsubtr}) provided that the pion vector form factor satisfies 
$F_\pi^V(s) \sim 1/s$ \cite{Lepage:1979zb}.

Let us come back to low energies and spell out the crucial approximation: 
The left-hand cut structure $K$ is determined from tree-level nucleon and (optionally) 
Delta exchange diagrams. 
This is essentially relativistic leading-order (LO) chiral perturbation theory ($\chi$PT) \cite{Becher:2001hv,Scherer:2012xha} 
with or without explicit Delta degrees of freedom \cite{Pascalutsa:2006up}. 
Extension
to next-to-leading order (NLO) does not provide additional diagrams with left-hand cuts. The effect can be encoded in the 
subtraction constant $P_0$. In other words one has
\begin{eqnarray}
  \label{eq:Tchipt}
  T^{\rm NLO\,\chi PT} = K + P_0  \,.
\end{eqnarray}
Note that a deviation of the Omn\`es function 
$\Omega$ from unity encodes pion rescattering and is therefore a loop effect in $\chi$PT. 
But NLO accuracy of baryon $\chi$PT means tree level \cite{Scherer:2012xha}. This justifies the result (\ref{eq:Tchipt}).
Complementary the constant $P_0$ can be determined from a fit to the radius \cite{Granados:2017cib}. I will come back to 
this aspect in section \ref{sec:chipt}.

\begin{figure}[ht] 
  \centering
  \begin{minipage}[c]{0.48\textwidth}  
    \includegraphics[keepaspectratio,width=\textwidth]{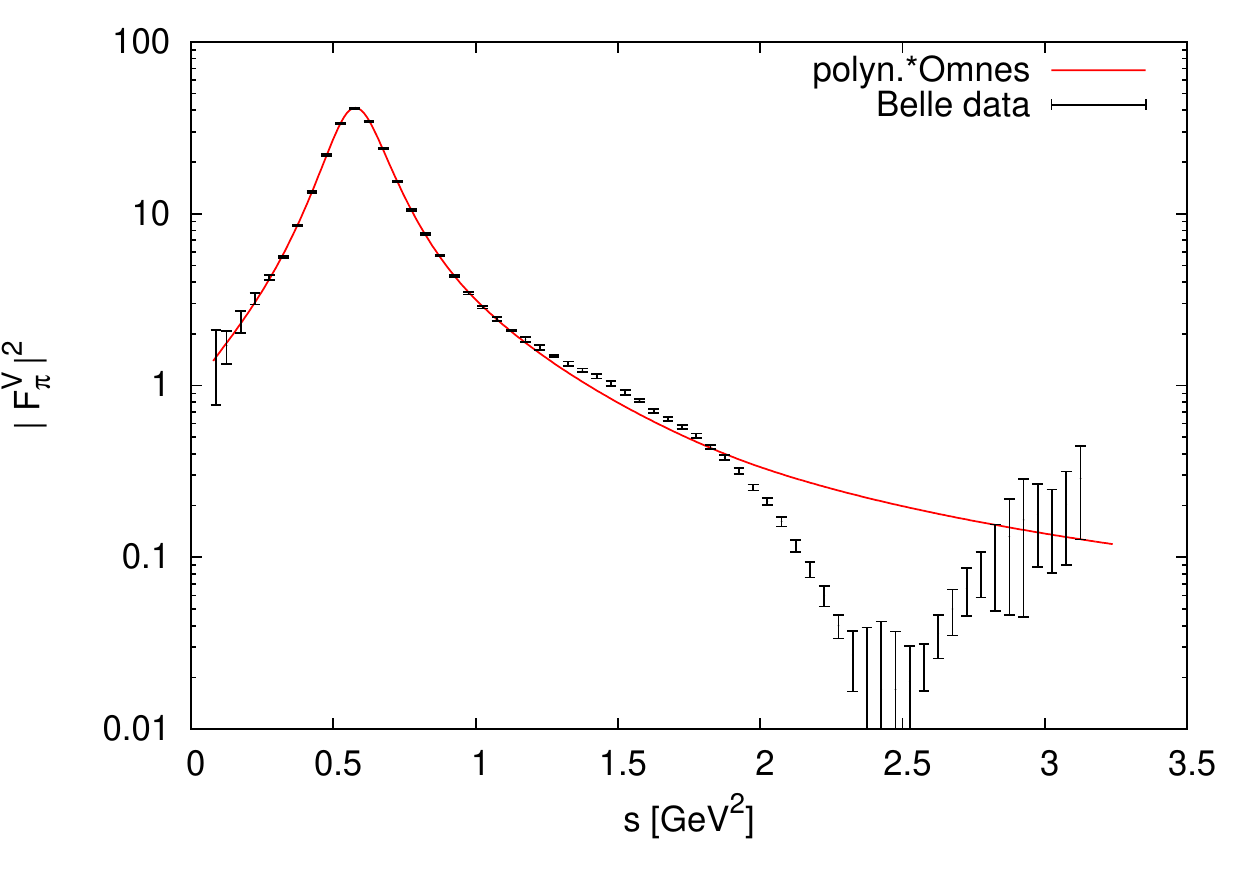}
  \end{minipage}   
  \caption{Pion vector form factor (modulus squared) from (\ref{eq:FV-Omnes-alphaV}) 
    as compared to Belle data \cite{Fujikawa:2008ma}.}
  \label{fig:pionVFF}
\end{figure}

The final ingredient is the pion vector form factor $F^V_\pi$ introduced in (\ref{eq:pionFFdef}). Here I slightly improve on the 
approximation of \cite{Granados:2017cib} and use \cite{Hanhart:2012wi,Hanhart:2013vba,Hoferichter:2016duk}
\begin{eqnarray}
  \label{eq:FV-Omnes-alphaV}
  F^V_\pi(s) = (1+\alpha_V \, s) \, \Omega(s) \,.
\end{eqnarray}
In practice I use the pion phase shift from \cite{GarciaMartin:2011cn} smoothly extrapolated to reach $\pi$ at 
infinity \cite{Hanhart:2012wi}. The parameter $\alpha_V$ is determined from a fit to data on the pion vector form factor from 
tau decays. A value of 
\begin{eqnarray}
  \label{eq:alphaV}
  \alpha_V = 0.12 \, {\rm GeV}^{-2}
\end{eqnarray}
yields the curve shown in figure \ref{fig:pionVFF}.
An excellent agreement is achieved for energies below 1 GeV. One should not expect a good agreement at higher energies where 
other intermediate states (four pions, six pions, \ldots) also play an important role.
Note that in contrast to \cite{Hanhart:2012wi,Hoferichter:2016duk} isospin breaking, in particular rho-omega mixing, 
is entirely ignored in the present work. 

I turn now to the second unitarization method. According to the optical theorem, at low energies the imaginary part of an 
isovector form factor is proportional to the product $T \, F_\pi^{V*}$, cf.\ (\ref{eq:dispbasic}), (\ref{eq:dispbasicunsubtr}). 
Following \cite{Alarcon:2017asr} this can be rewritten as 
\begin{eqnarray}
  \label{eq:NoverD-Weiss}
  T \, F_\pi^{V*} = \frac{T}{F_\pi^V} \, \vert F_\pi^V\vert^2 \approx 
  \left(\frac{T}{F_\pi^V}\right)_{\rm \chi PT} \; \left(\vert F_\pi^V\vert^2\right)_{\rm data}  \,.
\end{eqnarray}
The ratio of scattering amplitude and pion vector form factor is approximated by $\chi$PT. Right-hand cuts 
cancel out in this ratio. This construction resembles the N/D method. 
In \cite{Alarcon:2017asr} the modulus square of the pion vector form factor is taken from a fit to data. By construction it 
contains the correct right-hand cut. 

To compare this N/D approach to the MO scheme I use in the following $\chi$PT at NLO, i.e.\ equation (\ref{eq:Tchipt}) 
together with the phenomenologically
successful approximation (\ref{eq:FV-Omnes-alphaV}) for the pion vector form factor. The deviation of $F^V_\pi$ from unity 
encodes pion rescattering and is therefore a loop effect in $\chi$PT. 
In contrast, NLO accuracy of baryon $\chi$PT means tree level \cite{Scherer:2012xha}. Thus the N/D method 
of \cite{Alarcon:2017asr} yields at NLO:
\begin{eqnarray}
  \label{eq:NoverD-WeissX}
  \left(\frac{T}{F_\pi^V}\right)_{\rm \chi PT} \approx P_0 + K \,.
\end{eqnarray}
In contrast the MO scheme gives for the same quantity 
\begin{eqnarray}
  && \left(\frac{T}{F_\pi^V}\right)_{\rm MO} \approx  \frac{T^{\rm MO}}{\Omega} = \nonumber \\ 
  && P_0 + \frac{K}{\Omega} + s \, 
    \int\limits_{4m_\pi^2}^\infty \, \frac{ds'}{\pi} \, 
    \frac{\sin\delta(s') \, K(s')}{\vert\Omega(s')\vert \, (s'-s-i \epsilon) \, {s'}}  
  \label{eq:MO-comp}
\end{eqnarray}
where I have replaced a factor $(1+\alpha_V \, s)$ by unity. Below we will see that the results from MO and N/D deviate by 
factors of more than 2 in the region below 1 GeV. 
Thus effects on the 10\% level as caused by $\alpha_V$ can be safely neglected for the 
comparison. 

Inspecting the MO expression (\ref{eq:MO-comp}) and the N/D expression (\ref{eq:NoverD-WeissX}) we see that polynomials 
(here the constant $P_0$) are treated in the same way whereas left-hand cut structures are treated very differently. In general
\begin{eqnarray}
  K(s) & \not\approx & \frac{K(s)}{\Omega(s)} + s \, \int\limits_{4m_\pi^2}^\infty \, \frac{ds'}{\pi} \, 
  \frac{\sin\delta(s') \, K(s')}{\vert\Omega(s')\vert \, (s'-s-i \epsilon) \, {s'}} \nonumber \\
  & = & \frac{K(s)}{\vert \Omega(s) \vert} e^{-i\delta} + s \, \int\limits_{4m_\pi^2}^\infty \, \frac{ds'}{\pi} \, 
  \frac{\sin\delta(s') \, K(s')}{\vert\Omega(s')\vert \, (s'-s-i \epsilon) \, {s'}}  \nonumber \\ 
  & = & \frac{K(s)}{\Omega(s)} - s \, \int\limits_{4m_\pi^2}^\infty \, \frac{ds'}{\pi} \, 
  \frac{{\rm Im}\Omega^{-1}(s') \, K(s')}{(s'-s-i \epsilon) \, {s'}} \,.  
  \label{eq:non-eq}
\end{eqnarray}
I have provided several versions to display the MO structure to make sure that the analytic properties can be fully appreciated.
Note in particular that above the two-pion threshold --- the relevant region in the 
dispersive integrals (\ref{eq:dispbasic}), (\ref{eq:dispbasicunsubtr}) --- both the left- and the right-hand side of 
(\ref{eq:non-eq}) have no imaginary parts. Thus only the real parts differ. Essentially this ensures that Watson's theorem is 
satisfied for both approaches. Thus it is analyticity, not unitarity (the optical theorem) where the two approaches differ. 

Finally it should be stressed that in practice the dispersive 
integrals in (\ref{eq:dispbasic}), (\ref{eq:dispbasicunsubtr}) and (\ref{eq:tmandel}) are cut off at $\Lambda^2$. 
Similar to \cite{Granados:2017cib} I explore for $\Lambda$ the values 1 and 1.8 GeV. Note that cutoff values above 
the $N \bar N$ threshold would not be reasonable.

\section{Input from chiral perturbation theory}
\label{sec:chipt}

Before turning to the results I shall further specify the input of the calculations. Formally the same Lagrangians as 
in \cite{Granados:2017cib} are used. The pion-nucleon tree-level amplitudes can be obtained from the expressions given in 
\cite{Granados:2017cib} by the replacements $m_\Sigma, m_\Lambda \to m_N$, $m_{\Sigma^*} \to m_\Delta$, $D \, F/\sqrt{3} \to g_A^2/4$,
and $h_A^2/\sqrt{3} \to 2 h_A^2/3$ 
with $g_A = F+D = 1.26$; see also appendix \ref{sec:app0}. 
The pion-Delta-nucleon coupling constant $h_A$ is chosen such that 
the width of the Delta is reproduced. I use $h_A = 2.88$. Note that the value obtained from 
hyperon decays and used in \cite{Granados:2017cib} is significantly smaller ($h_A \approx 2.3$). However, the whole framework 
that highlights the dominant role of the light pions and disregards kaons is not SU(3) symmetric anyway; see also 
the corresponding discussion in \cite{Granados:2017cib}. Thus it appears most appropriate that the three-point coupling constants 
that are required in exchange diagrams are determined from corresponding two-body decay widths. This is the phenomenology-based
philosophy followed here and in \cite{Granados:2017cib}. Only in the absence of phenomenological input flavor SU(3) is utilized 
as a fall-back option. 

Following \cite{Granados:2017cib} I will discuss in the following three approximations for the $\chi$PT input:
\begin{enumerate}
\item ``LO'': Purely-nucleon $\chi$PT at LO, i.e.\ no explicit Delta degrees of freedom; 
  essentially these are the Born diagrams of nucleon exchange and a contact interaction 
  from the Weinberg-Tomozawa term \cite{Weinberg:1966kf,Tomozawa:1966jm}.
\item ``NLO'': Purely-nucleon $\chi$PT at NLO; as pointed out in \cite{Granados:2017cib} there is no modification for the 
  electric case
  while for the magnetic case there is a contribution from a contact interaction of the NLO Lagrangian. 
  In the SU(2) $\chi$PT language of \cite{Fettes:1998ud,Becher:2001hv} this term is proportional to the low-energy constant $c_4$.
\item ``NLO+res'': Nucleon+Delta $\chi$PT at NLO; the overall strength of the contact interaction is adjusted 
  such that at low energies
  the result matches to the previous case, see \cite{Granados:2017cib} for details.
\end{enumerate}
With the third case one can study to which extent a dynamical treatment of the Delta really matters.

For the N/D method it is not necessary to specify $K$ and $P_0$ separately. According to (\ref{eq:NoverD-WeissX}) only the 
sum matters. However, in the MO formula (\ref{eq:tmandel}) the two ingredients $P_0$ and $K$ are treated differently.
One can consider what happens if a constant is kept as part of $K$. 
This can be deduced from the last expression in (\ref{eq:non-eq}) and is discussed in more detail in appendix \ref{sec:app1}; 
see also the corresponding discussion in \cite{Kang:2013jaa}. The result of these considerations is that (\ref{eq:tmandel}) 
would be modified. Actually the additional term changes the high-energy behavior.
To keep the appealing high-energy behavior described after (\ref{eq:tmandel}) it is of advantage to keep a $K$ that vanishes 
at large energies apart from the constant $P_0$. 

Thus it is necessary to specify $K$ and $P_0$ 
separately. In addition, one might adopt a point of view that is somewhat 
complementary to chiral perturbation theory. The importance of nucleon and Delta exchange as the most relevant left-hand cuts 
can also be motivated on phenomenological grounds. What remains to be determined is then the constant $P_0$, to be more specific:
one constant for the electric and one for the magnetic sector. This can be achieved by a fit to the corresponding radius. 
Thus it makes sense to fully specify $K$ and $P_0$ separately. 

In general, a calculation of tree-level nucleon and Delta exchange diagrams yields polynomials and left-hand cut
structures that cannot be further reduced by partial fraction decomposition. Only the latter are subsumed in $K$. As discussed 
in \cite{Granados:2017cib}, $K$ does not depend on the chosen representation for the fields while in general the polynomial does.
One might dub this ``offshell ambiguity''. In an effective field theory this ambiguity is compensated by the appearance of 
contact interactions \cite{Fearing:1999fw}. 

For the case considered here, $K$ consists of contributions from nucleon and from Delta exchange. I call these contributions 
$K^{\rm Born}$ and $K^{\rm res}$, respectively. Correspondingly I introduce
\begin{eqnarray}
  && T^{\rm Born/res}(s)  :=  \nonumber \\ && K^{\rm Born/res}(s) + \Omega(s) \, s \, 
    \int\limits_{4m_\pi^2}^\infty \, \frac{ds'}{\pi} \, 
    \frac{\sin\delta(s') \, K^{\rm Born/res}(s')}{\vert\Omega(s')\vert \, (s'-s-i \epsilon) \, {s'}} \,. \nonumber \\ && 
  \label{eq:tmandelsep}
\end{eqnarray}
Then the MO scattering amplitude (\ref{eq:tmandel}) can be written as 
\begin{eqnarray}
  \label{eq:TMObornrespoly}
  T^{\rm MO}(s) = T^{\rm Born}(s) + T^{\rm res}(s) + \Omega(s) \, P_0  \,.
\end{eqnarray}

Instead of determining $P_0$ from $\chi$PT one might fit it to the radius. Using (\ref{eq:dispbasicradii})
this reads for the MO scheme:
\begin{eqnarray}
  \langle r^2_{M/E} \rangle & = &
  \int\limits_{4 m_\pi^2}^\infty \frac{ds}{\pi} \, 
  \frac{\left(T^{\rm Born}_{M/E}(s) + T^{\rm res}_{M/E}(s) \right)\, p_{\rm c.m.}^3(s) \, F^{V*}_\pi(s)}{2 \pi \, s^{5/2}}  
  \nonumber \\ && {}
  + P_{0,M/E} \, \int\limits_{4 m_\pi^2}^\infty \frac{ds}{\pi} \, 
  \frac{\Omega(s) \, p_{\rm c.m.}^3(s) \, F^{V*}_\pi(s)}{2\pi \, s^{5/2}}  \,.
  \label{eq:dispbasicradiifitP0}  
\end{eqnarray}

In the magnetic sector there is a contribution to $P_{0,M}$ from the NLO Lagrangian. It is proportional to the low-energy constant 
$c_4$. Here one might turn the line of reasoning around and determine $c_4$ from the isovector magnetic radius of the 
nucleon. 

The explicit expressions for $K$ can be easily obtained from the formulae given in \cite{Granados:2017cib} together with the 
replacement rules specified in the beginning of this section. The $\chi$PT expressions for $P_0$ are as follows. At LO of 
purely-nucleon $\chi$PT one obtains
\begin{eqnarray}
  \label{eq:P0LO}
  P^{\rm LO}_{0,M} = P^{\rm LO}_{0,E} = - \frac{g_A^2-1}{2 \, F_\pi^2}  \,.
\end{eqnarray}
At NLO of purely-nucleon $\chi$PT one finds the additional contribution
\begin{eqnarray}
  \label{eq:P0NLO}
  P^{\rm NLO}_{0,M} = \frac{2 \, m_N \, c_4}{F_\pi^2}  \,.
\end{eqnarray}
In the electric sector there is no NLO modification.

In the ``NLO+res'' approximation that includes dynamical Delta baryons the additional contributions to 
$P_0$ depend on the representation for the 
Delta fields \cite{Granados:2017cib}. In the electric sector this ambiguity is relegated to higher orders. One gets
\begin{eqnarray}
  \label{eq:P0resE}
  P^{\rm res}_{0,E} = \frac{h_A^2 \, (m_N+m_\Delta)^2}{36 \, m_\Delta^2 \, F_\pi^2} \,.
\end{eqnarray}
In the magnetic sector this ambiguity persists, but is compensated by the appearance of the NLO contact term $\sim c_4$. 
It makes sense to demand that the same low-energy limit is obtained in the effective theories with and without the 
Delta resonance. 
In the magnetic sector this requires a modification of $c_4$. Alternatively one can keep the value of 
$c_4$ and subtract the contribution from $K^{\rm res}_M(s)$ evaluated at an appropriate low-energy point $s$. 
Following the procedure 
outlined in \cite{Granados:2017cib} I choose
\begin{eqnarray}
  P^{\rm res}_{0,M} & = & - \lim_{s \to 0} \lim_{m_\pi \to 0} K^{\rm res}_M(s) \nonumber \\ 
  & = & -\frac{h_A^2 \, (4 m_\Delta \, m_N - m_\Delta^2 - m_N^2) \, (m_\Delta + m_N)}{36 \, m_\Delta^2 \, (m_\Delta - m_N) \, F_\pi^2} \,.
  \phantom{mm}
  \label{eq:P0resM}
\end{eqnarray}

Purely-nucleon $\chi$PT at LO means to consider only (\ref{eq:P0LO}). Purely-nucleon $\chi$PT at NLO means to sum for the 
magnetic sector (\ref{eq:P0LO}) and (\ref{eq:P0NLO}). Nucleon+Delta  $\chi$PT at NLO (``NLO+res'') means to sum all contributions 
(\ref{eq:P0LO})-(\ref{eq:P0resM}). 

\section{Results}
\label{sec:results}

\subsection{Magnetic sector}
\label{sec:submag}

The deviations in the treatment of left-hand cut structures when comparing N/D and MO 
suggest that significant quantitative differences might appear for left-hand cuts that start close to the threshold of the 
right-hand cut, the two-pion threshold. Indeed the nucleon exchange diagrams provide a pole at 
$s= 4 m_\pi^2 - \frac{m_\pi^4}{m_N^2}$ \cite{Frazer:1960zzb}, i.e.\ very close to threshold. As I will show below, 
significant differences between the results of the two methods appear. 

One might guess that the MO method 
should yield more reliable results since it decomposes thoroughly the analytic structure of the amplitudes. On the other hand,
it must be stressed that for both methods the input comes from $\chi$PT and not directly from data. It is not 
guaranteed a priori which combination of input and unitarization method is capable to provide the most reliable results.
For the case at hand the quality assessment will be provided by a comparison to the results from a fully dispersive 
analysis of pion-nucleon scattering based on Roy-Steiner (RS) equations \cite{Hoferichter:2015hva}. I will compare the imaginary 
parts of the nucleon form factors in the region between the two-pion threshold and 1 GeV 
as obtained from the MO scheme, 
\begin{eqnarray}
  \label{eq:MOImG}
  {\rm Im}G^{\rm MO}(s) = \frac{p_{\rm c.m.}^3}{12\pi \, \sqrt{s}} \, T^{\rm MO}(s) \, (1+\alpha_V \, s) \, \Omega^*(s) \,,
\end{eqnarray}
the N/D scheme, 
\begin{eqnarray}
  {\rm Im}G^{\rm N/D}(s) = \frac{p_{\rm c.m.}^3}{12\pi \, \sqrt{s}} \, \left( P_0 + K(s) \right) \, 
  \left\vert (1+\alpha_V \, s) \, \Omega(s) \right\vert^2  \,, 
  \nonumber \\ 
  \label{eq:NDImG}
\end{eqnarray}
and the RS analysis \cite{Hoferichter:2016duk}. 

\begin{figure}[ht]
  \centering
  \begin{minipage}[c]{0.48\textwidth}  
    \includegraphics[keepaspectratio,width=\textwidth]{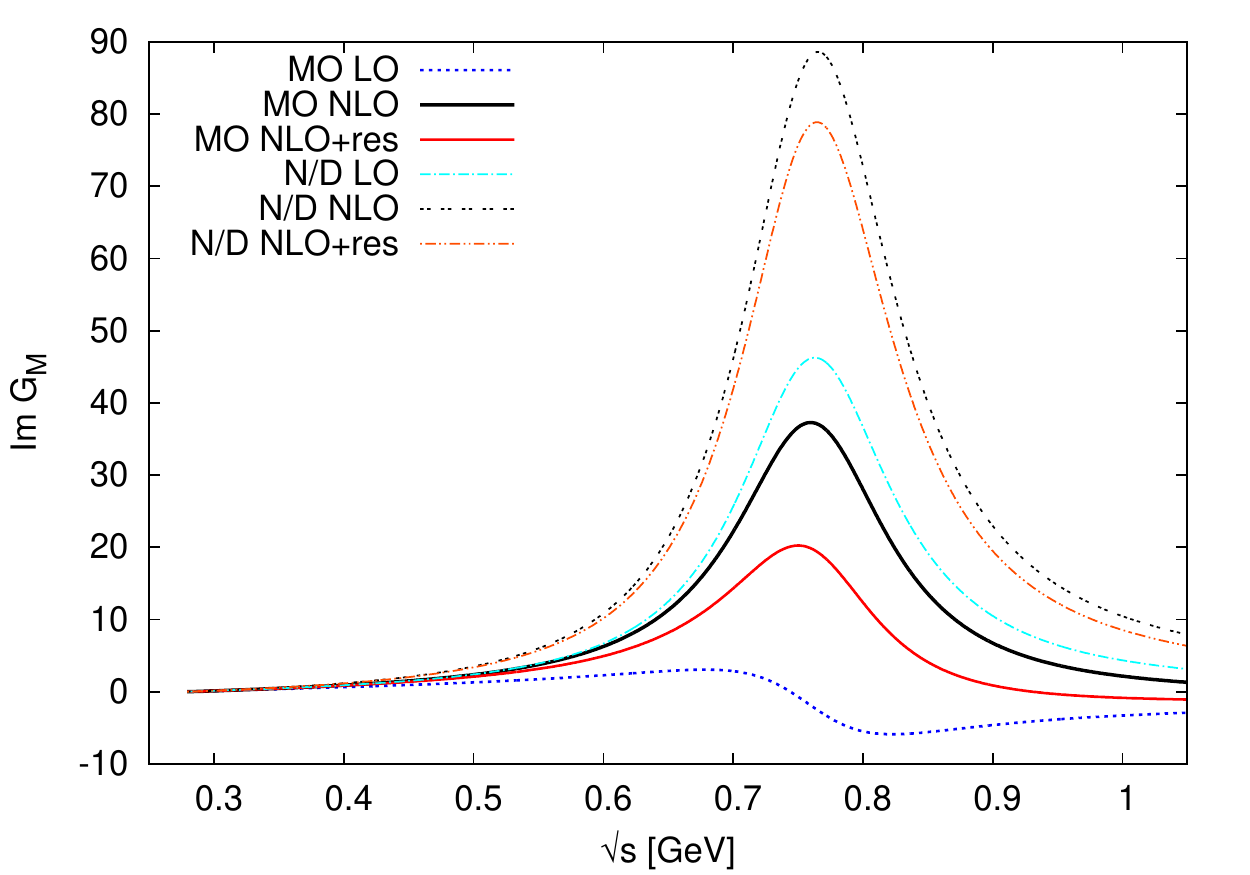}
  \end{minipage}   
  \caption{The imaginary part of the magnetic (isovector) form factor using various approximations: 
    ``MO'' refers to (\ref{eq:MOImG}),
    ``N/D'' to (\ref{eq:NDImG}). the labels ``LO'', ``NLO'' and ``NLO+res'' refer to the list of successive approximations
    specified in section \ref{sec:chipt}. For the non-color version of this figure the labels are assigned to the line ordering:
    ``MO LO'': bottom line; ``MO NLO'': third line from bottom; ``MO NLO+res'': second line from bottom;
    ``N/D LO''. third line from top; ``N/D NLO'': top line; ``N/D NLO+res'': second line from top. 
    $\Lambda=1.8 \,$GeV is used and
    the constant $P_0$ has been chosen such that the magnetic radius is reproduced by MO NLO+res.}
  \label{fig:compMOND}
\end{figure}

The imaginary part of the magnetic form factor is shown in figure \ref{fig:compMOND} for MO and N/D using the same input.
Obviously the results differ substantially in the rho-meson region. The constant $P_0$ is adjusted such that the magnetic radius 
used in \cite{Hoferichter:2016duk} is reproduced in the MO scheme using nucleon and Delta 
exchange.\footnote{It is worth to mention that the value for the isovector magnetic radius used in  \cite{Hoferichter:2016duk} 
deviates quite a bit from the value extracted from the data collected in \cite{pdg}; see also the corresponding 
discussion in \cite{Hoferichter:2016duk}. Since I use the radius as an input, not as 
an output, there is no point in a detailed investigation of this disagreement for the present work.}
Translated to purely-nucleon $\chi$PT \cite{Fettes:1998ud} this corresponds to an NLO 
low-energy constant $c_4 = 2.99\,$GeV$^{-1}$. The low-energy constants of pion-nucleon scattering have been 
determined in \cite{Hoferichter:2015tha,Hoferichter:2015hva} by matching the dispersive RS representation to 
the $\chi$PT representation in the sub-threshold region. The results for $c_4$ are: $c_4^{\rm NLO} \approx 2.17 \,$GeV$^{-1}$ 
and $c_4^{\rm NNLO} \approx 3.56 \,$GeV$^{-1}$. 
Given that the MO scheme goes beyond NLO $\chi$PT by including pion rescattering
but does not provide a full one-loop $\chi$PT calculation, it should be expected that the value of $c_4$ lies between the 
values from NLO and NNLO. This is indeed the case. If the cutoff $\Lambda$ is changed from 1.8 GeV to 1 GeV, then a slight 
readjustment of the value for $c_4$ is required to reproduce the same value for the magnetic radius, 
now $c_4 = 3.07\,$GeV$^{-1}$. 

\begin{figure}[ht]
  \centering
  \begin{minipage}[c]{0.48\textwidth}  
    \includegraphics[keepaspectratio,width=\textwidth]{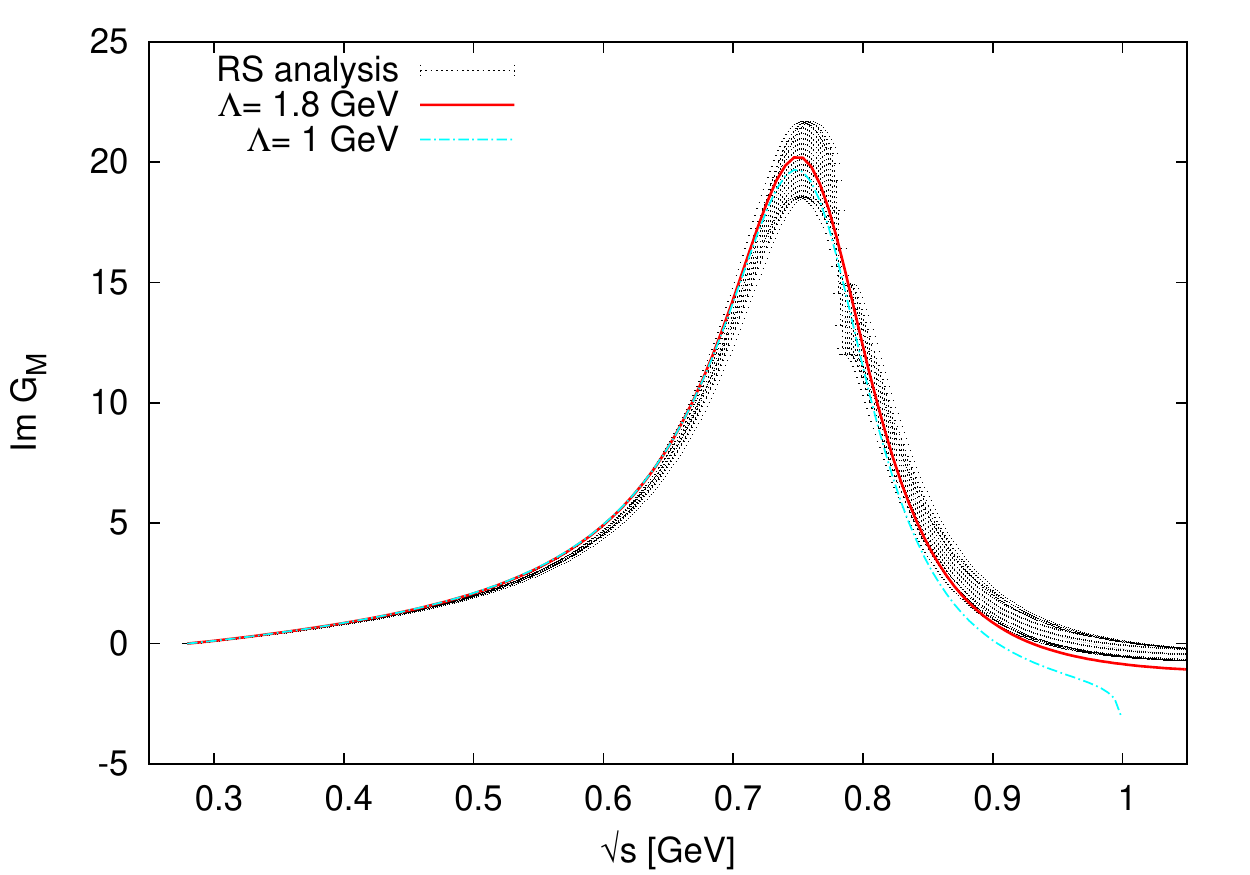}
  \end{minipage}   
  \caption{The imaginary part of the magnetic (isovector) form factor using the MO NLO+res scheme for two values of the 
    cutoff $\Lambda$. For both cutoffs the 
    respective constant $P_{0,M}$ has been chosen such that the magnetic radius is reproduced. 
    The band denotes the results from the RS analysis. Note that the full red lines here and in figure \ref{fig:compMOND} show 
    the same result.}
  \label{fig:compMORS}
\end{figure}

\begin{figure}[ht]
  \centering
  \begin{minipage}[c]{0.48\textwidth}  
    \includegraphics[keepaspectratio,width=\textwidth]{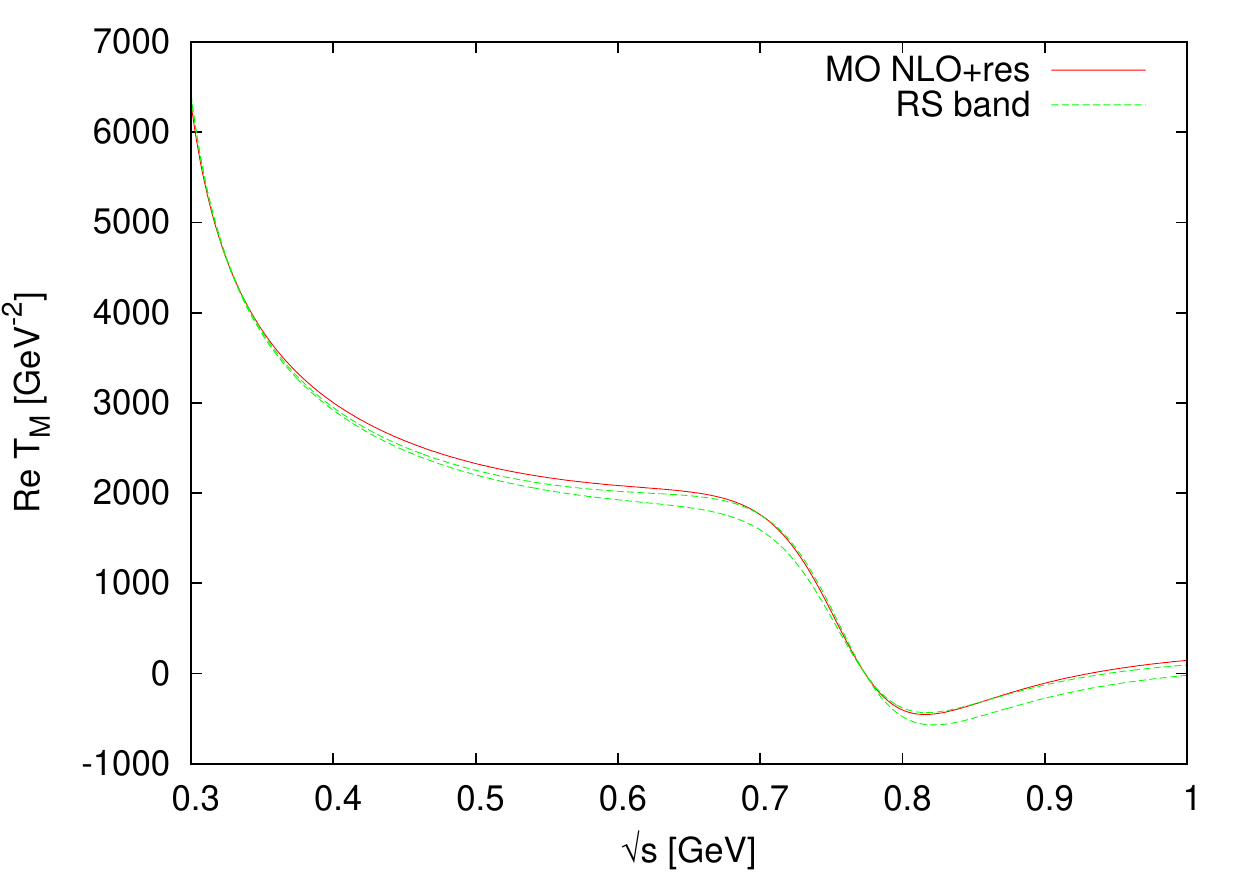}

    \includegraphics[keepaspectratio,width=\textwidth]{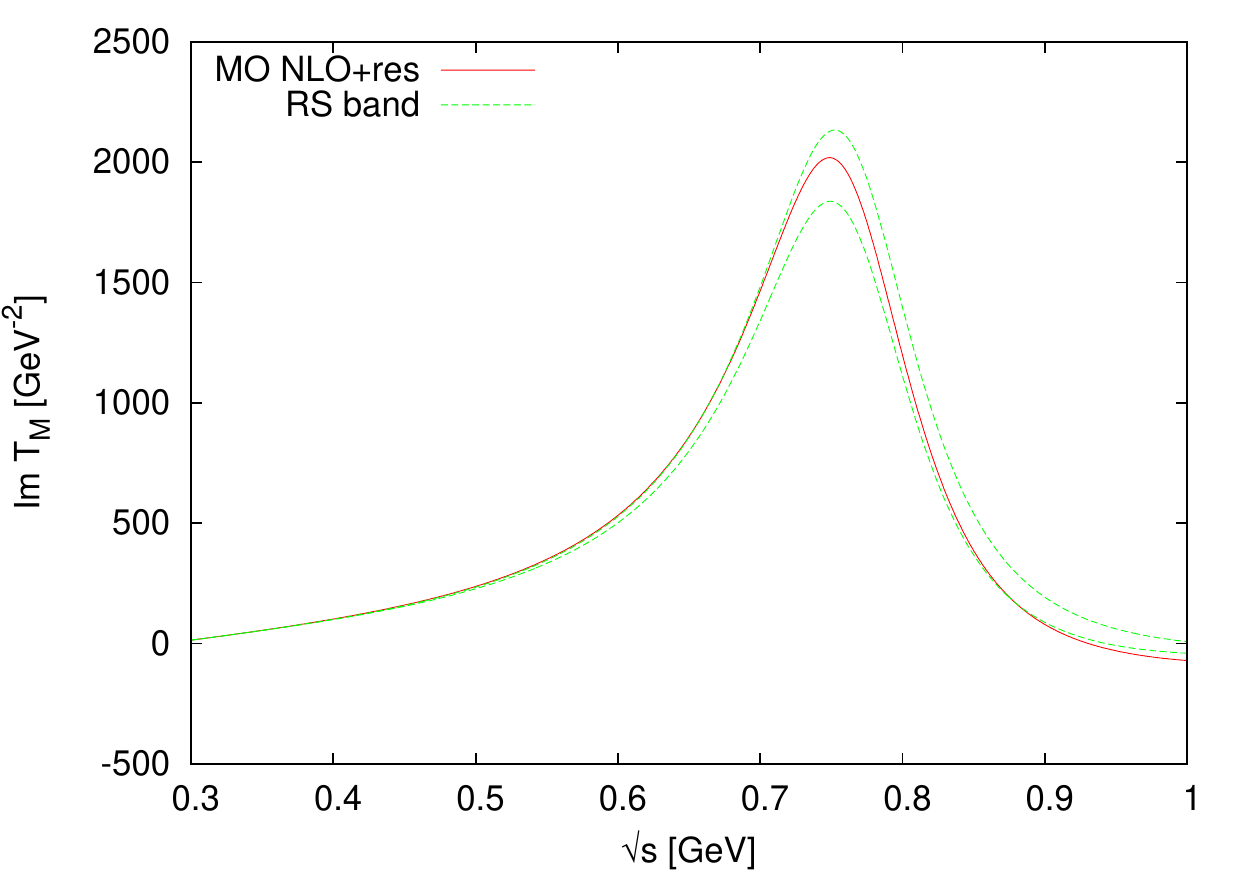}
  \end{minipage}   
  \caption{Real (top) and imaginary (bottom) part of the magnetic amplitude comparing RS  \cite{Hoferichter:2015hva}
    and MO NLO+res. $\Lambda = 1.8\,$GeV
    and $c_4  = 2.99\,$GeV$^{-1}$ have been used.}
  \label{fig:compReImT}
\end{figure}

The MO results using nucleon+Delta $\chi$PT at NLO for two different cutoffs $\Lambda$ are compared to the RS result 
of \cite{Hoferichter:2016duk} in figure \ref{fig:compMORS}. 
Obviously excellent agreement is obtained between MO and RS up to the region where differences caused by different cutoffs 
matter. This is only beyond the rho-meson peak region. With the same input the N/D result deviates already in the rho-meson 
region significantly as shown in figure \ref{fig:compMOND}. If one tried to obtain a spectrum comparable to the ones of 
figure \ref{fig:compMORS} using N/D NLO+res, then one would need $c_4 \approx -1.1\,$GeV$^{-1}$, a value that differs in size 
and sign from the values extracted in \cite{Hoferichter:2015tha,Hoferichter:2015hva}. With 
realistic values for the low-energy constant $c_4$ the N/D spectrum is much larger than the RS spectrum in the 
rho-meson region. Note that this is exactly what has been found in \cite{Alarcon:2017asr}; see figure 7 therein. 
In \cite{Alarcon:2017asr} the reasonable agreement between N/D and RS {\em below} the rho-peak has been stressed. 
it has been argued that the disagreement in the rho-meson region is less important for the {\em peripheral} transverse densities.
However, the interesting point is that a better agreement over a larger range can be achieved by the use of the 
MO method instead of N/D. 
To substantiate this further I compare in figure \ref{fig:compReImT} directly 
the magnetic amplitude for MO and RS \cite{Hoferichter:2015hva}. Again impressive agreement is achieved. 

From now on I focus on the MO scheme. Returning to figure \ref{fig:compMOND} one observes that the inclusion of NLO and of 
dynamical Deltas both matters for the spectral information. To see whether this also matters for the low-energy quantities 
I turn now to the determination of magnetic moment and radius based on (\ref{eq:dispbasicunsubtrkappa}) 
and (\ref{eq:dispbasicradii}), respectively. Results are shown in table \ref{tab:mag} for two values of the cutoff $\Lambda$.
\begin{table}[h!]
  \centering
  \begin{tabular}{|l|r|r|}
    \hline
    $G_M(0)$ & $\Lambda = 1\,$GeV & $\Lambda = 1.8\,$GeV \\ \hline
    LO & $-0.37$ & $-0.72$  \\ \hline
    NLO & 5.94 & 6.12  \\ \hline
    NLO+res & 3.19 & 3.03  \\ \hline
    exp. & \multicolumn{2}{|c|}{2.35}    \\ \hline
    \hline 
    $\langle r^2_M \rangle$ [GeV$^{-2}$] & $\Lambda = 1\,$GeV & $\Lambda = 1.8\,$GeV \\ \hline
    LO & 6.30 & 6.52  \\ \hline
    NLO & 75.30 & 76.72  \\ \hline
    NLO+res & 46.73 & 46.79  \\ \hline
    experiment & \multicolumn{2}{|c|}{46.76}   \\ \hline
  \end{tabular}
  \caption{Results for the isovector magnetic moment and radius for two different cutoffs. Data are taken from \cite{pdg} for 
    the magnetic moment and from \cite{Hoferichter:2016duk} for the radius.}
  \label{tab:mag}
\end{table}
Note that the NLO low-energy constant $c_4$ has always been chosen such that for NLO+res the correct radius is obtained.
As already pointed out the obtained values for $c_4$ are very realistic and lead to scattering amplitudes that agree with the 
RS results (figure \ref{fig:compReImT}).

One observes that LO alone, i.e. Born diagrams and Weinberg-Tomozawa term, does not provide realistic values. With the inclusion
of the NLO term the correct orders of magnitude for both quantities, magnetic moment and radius, are achieved. The inclusion 
of dynamical Deltas has also a non-negligible quantitative impact. It should not be surprising that the magnetic moment is 
not fully reproduced. The unsubtracted dispersion relation (\ref{eq:dispbasicunsubtrkappa}) is too sensitive to the high-energy
part of the integrand, which is not fully under control; see also the discussion in \cite{Hoferichter:2016duk}. The dependence on 
the cutoff $\Lambda$ is of minor importance, a reassuring result given that a low-energy theory is used. In principle, one 
could also study the impact of changes in $h_A$ and $g_A$ and in the pion phase shift. The results would not change qualitatively 
and the low-energy constant $c_4$ can always be readjusted to obtain the radius in the full NLO+res approximation. For the 
electric sector I will study the impact of a variation in the pion-Delta-nucleon coupling constant $h_A$ 
in subsection \ref{sec:subel}. To summarize, I find the very same pattern as for the hyperon case discussed 
in \cite{Granados:2017cib} giving further credit to the ideas spelled out there.

\begin{figure}[ht]
  \centering
  \begin{minipage}[c]{0.48\textwidth}  
    \includegraphics[keepaspectratio,width=\textwidth]{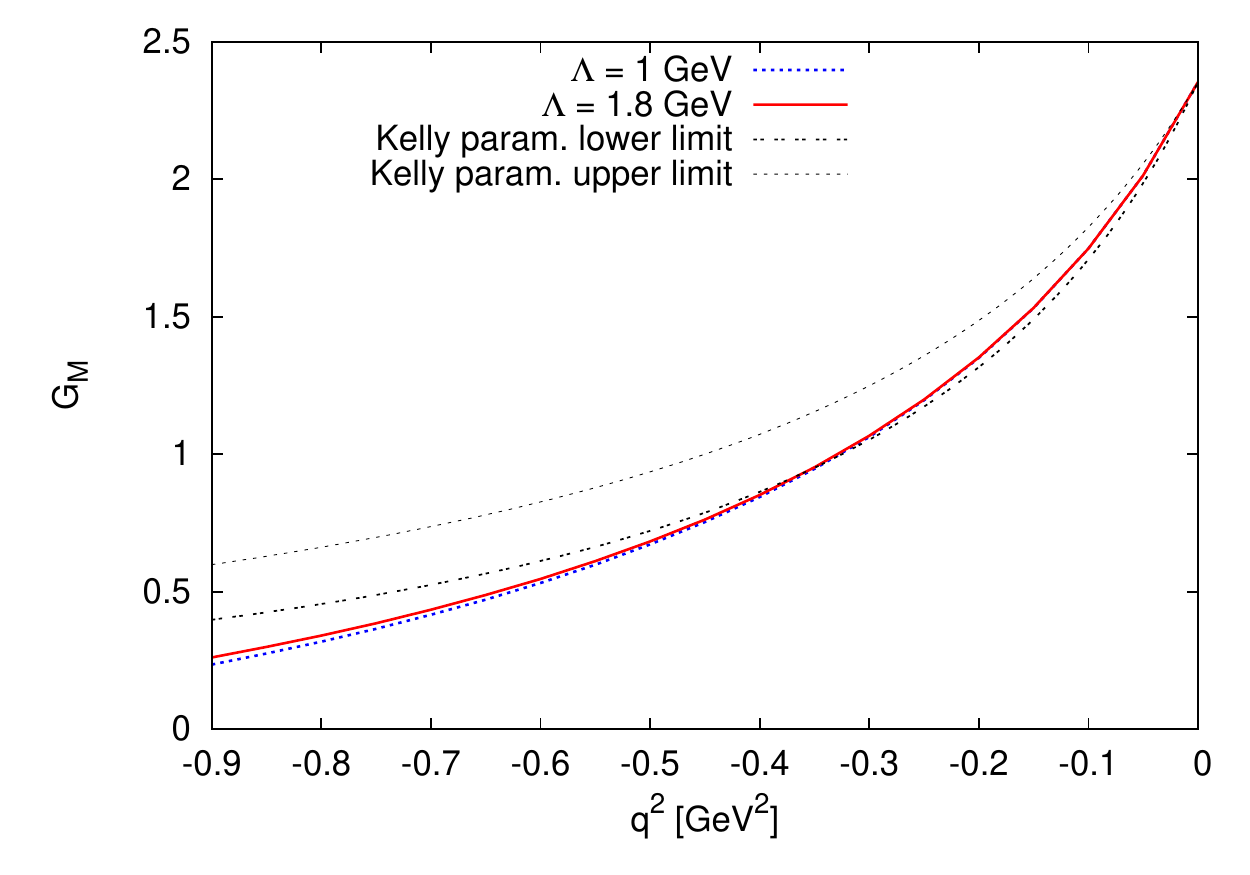}
  \end{minipage}   
  \caption{The (isovector part of the) magnetic form factor of the nucleon in the spacelike region using two different values of
    the cutoff $\Lambda$ as compared to the Kelly parametrization \cite{Kelly:2004hm}.}
  \label{fig:GMspace}
\end{figure}
In figure \ref{fig:GMspace} the (isovector part of the) magnetic form factor is compared to the 
Kelly parametrization \cite{Kelly:2004hm} . The latter was obtained from a fit to proton and neutron data. 
The calculations are based on the subtracted dispersion relation (\ref{eq:dispbasic}) using the MO scheme with nucleon and 
Delta exchange, i..e.\ NLO+res. 
Variation of the cutoff turns out to be fairly irrelevant. 
Note that the agreement of value and slope at the photon point is by construction (adjusting $G_M(0)$ and $P_{0,M}$), but the 
reasonable agreement between the results and the Kelly parametrization extends much further than a pure agreement of the 
slope would provide. The agreement between the low-energy calculations of the present work and the Kelly parametrization up to 
$\vert q^2 \vert \approx 0.4\,$GeV$^2$ is quite encouraging. A direct comparison to proton and neutron data is hampered by 
the lack of the isoscalar part of the nucleon form factors. I leave this part to future investigations.

\subsection{Electric sector}
\label{sec:subel}

In the magnetic sector the combination of dispersion theory and input from nucleon+Delta $\chi$PT at NLO seems to work very well,
albeit one should be aware that possible shortcomings of the $\chi$PT input might be hidden by the low-energy constant $c_4$, 
which is to some extent adjustable. This is not possible in the electric sector where LO and NLO agree 
(note that I always use physical values for the nucleon mass and for $g_A$). In the electric sector only the explicit 
inclusion of dynamical Delta degrees of freedom makes a change.
Yet, (N)LO+res does not provide a satisfying value for the electric radius, as can be read off from table \ref{tab:el1}.
\begin{table}[h!]
  \centering
  \begin{tabular}{|l|r|r|}
    \hline
    $G_E(0)$ & $h_A = 2.88$ & $h_A = 2.67$ \\ \hline
    LO & \multicolumn{2}{|c|}{$-0.56$}   \\ \hline
    NLO+res & 0.39 & 0.26  \\ \hline
    exp. & \multicolumn{2}{|c|}{1/2}    \\ \hline
    \hline 
    $\langle r^2_E \rangle$ [GeV$^{-2}$] & $h_A = 2.88$ & $h_A = 2.67$ \\ \hline
    LO & \multicolumn{2}{|c|}{0.29}  \\ \hline
    NLO+res & 7.91 & 6.84  \\ \hline
    experiment & \multicolumn{2}{|c|}{11.02}   \\ \hline
  \end{tabular}
  \caption{Results for the isovector charge and isovector electric radius for two different values of the 
    pion-Delta-nucleon coupling
    constant $h_A$. $\Lambda = 1.8\,$GeV is used. Data are taken from \cite{pdg}.}
  \label{tab:el1}
\end{table}

Before inspecting possible shortcomings of $\chi$PT I will discuss the results of table \ref{tab:el1} in more detail.
Like for the magnetic case, LO alone provides values for (isovector) charge and radius that are off by an order of magnitude 
(or even sign). The inclusion of dynamical Deltas delivers the correct order of magnitude, but not an accurate value for the 
electric radius. I note in passing that the much smaller disagreement in the electric radius of the proton as extracted from 
electronic versus muonic hydrogen \cite{Pohl:2010zza,Carlson:2015jba} is of no concern for the present discussion.
Besides presenting the results for LO and NLO+res (which coincides with LO+res) I have also explored the impact of a 
variation in $h_A$. The choice $h_A = 2.88$ reproduces the width of the Delta baryon in a tree-level calculation. This evaluation
is consistent with the use of $h_A$ in the tree-level exchange diagrams. The choice 
\begin{eqnarray}
  \label{eq:hagalnc}
  h_A = 3 g_A/\sqrt{2} \approx 2.67  
\end{eqnarray}
is the value obtained for QCD in the limit of a large number of colors, $N_c$ \cite{Dashen:1993as,Granados:2017cib}. 
Table \ref{tab:el1} shows that a variation of $h_A$ in a reasonable range 
does not reproduce the isovector electric radius. The same is true for a variation in the pion-nucleon coupling constant $g_A$ 
(not shown here).

It is worth to inspect the LO=NLO nucleon+Delta $\chi$PT input in more detail. 
Actually there are large cancelation effects in the electric sector \cite{Granados:2013moa} 
that might cause a sensitivity to unaccounted higher-order terms. I note in passing that this cancelation does not happen in the
magnetic sector. From a formal point of view the cancelation can be best understood in the 
large-$N_c$ limit \cite{Dashen:1993as,Lam:1997ry}. In this limit 
the masses of Delta and nucleon are degenerate and the pion-baryon three-point coupling constants are related by 
(\ref{eq:hagalnc}). As already demonstrated in \cite{Granados:2017cib}, appendix A, the left-hand cut structures $K_E^{\rm Born}$ 
and $K_E^{\rm res}$ completely cancel each other in this limit. A significant part of this cancelation survives in the real world 
of $N_c=3$. It is illuminating to discuss this cancelation effect also for $P_{0,E}$. 
According to (\ref{eq:P0LO}), (\ref{eq:P0resE}) there are three terms in LO=NLO nucleon+Delta $\chi$PT originating from the 
Born diagrams,
\begin{eqnarray} 
  \label{eq:polyelborn}
  P^{\rm Born}_{0,E} := - \frac{g_A^2}{2 \, F_\pi^2} \approx - 93. \, {\rm GeV}^{-2} \quad \sim N_c  \,,
\end{eqnarray}
from the Weinberg-Tomozawa term,
\begin{eqnarray}
  \label{eq:P0WT}
  P^{\rm WT}_{0,E} :=  \frac{1}{2 \, F_\pi^2} \approx 59. \, {\rm GeV}^{-2} \quad \sim \frac{1}{N_c}  \,,
\end{eqnarray}
and from the Delta-resonance exchange diagrams
\begin{eqnarray}
  \label{eq:P0resE2}
  P^{\rm res}_{0,E} = \frac{h_A^2 \, (m_N+m_\Delta)^2}{36 \, m_\Delta^2 \, F_\pi^2} \approx 84. \, {\rm GeV}^{-2}  \quad \sim N_c  \,.
\end{eqnarray}
I have provided numerical values for the real world of three colors and the information how the terms scale with the 
number of colors. Using $m_\Delta \to m_N$ and (\ref{eq:hagalnc}) it is easy to check that in the large-$N_c$ limit 
the contributions (\ref{eq:polyelborn}) and (\ref{eq:P0resE2}) exactly cancel. Formally these contributions are separately of 
order $N_c$ and therefore much
larger than the remaining Weinberg-Tomozawa contribution, which is $N_c$ suppressed. By inspecting the numerical values in 
(\ref{eq:polyelborn}), (\ref{eq:P0WT}), (\ref{eq:P0resE2}) this ordering can still be seen qualitatively for $N_c=3$, albeit 
the Weinberg-Tomozawa contribution is not the order of magnitude smaller that a factor $1/N_c^2$ might suggest.

\begin{figure}[ht]
  \centering
  \begin{minipage}[c]{0.48\textwidth}  
    \includegraphics[keepaspectratio,width=\textwidth]{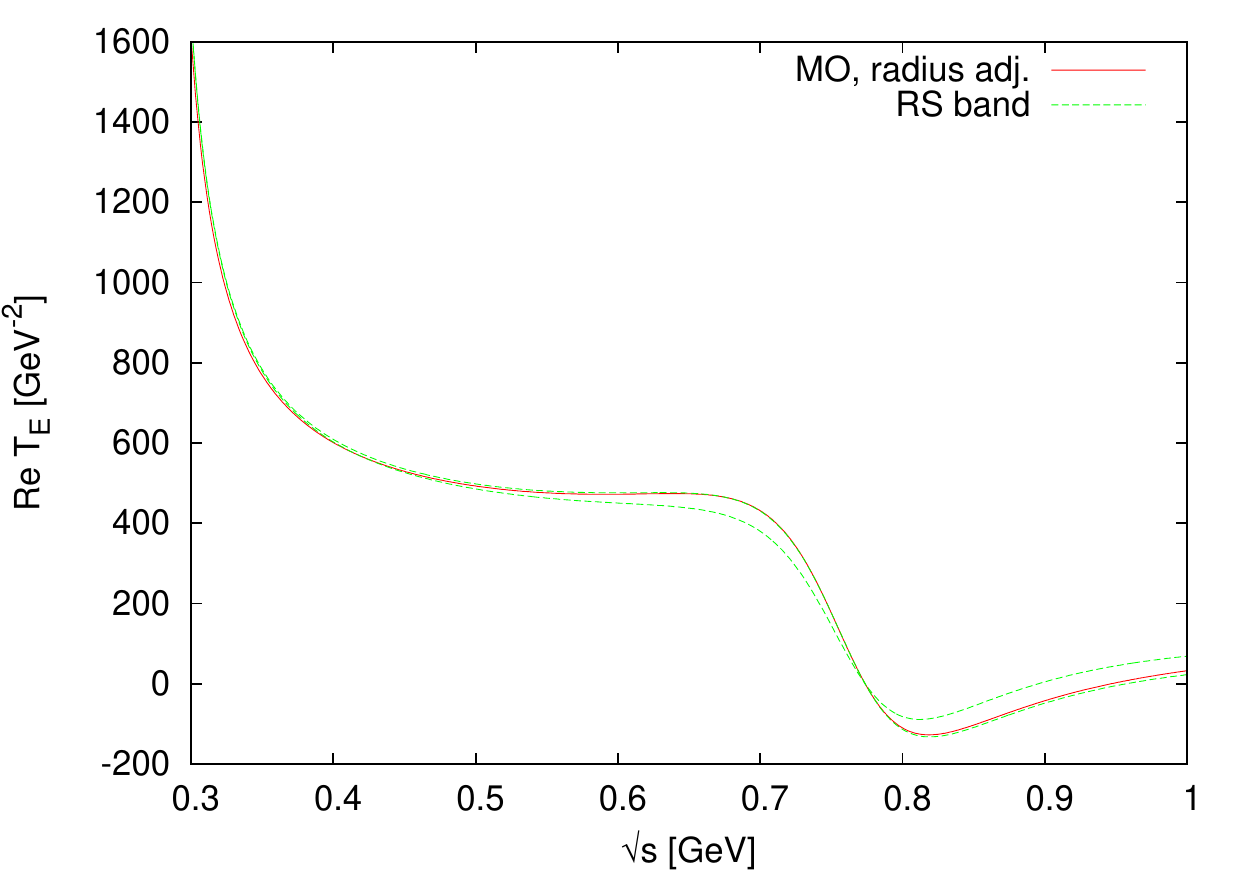}

    \includegraphics[keepaspectratio,width=\textwidth]{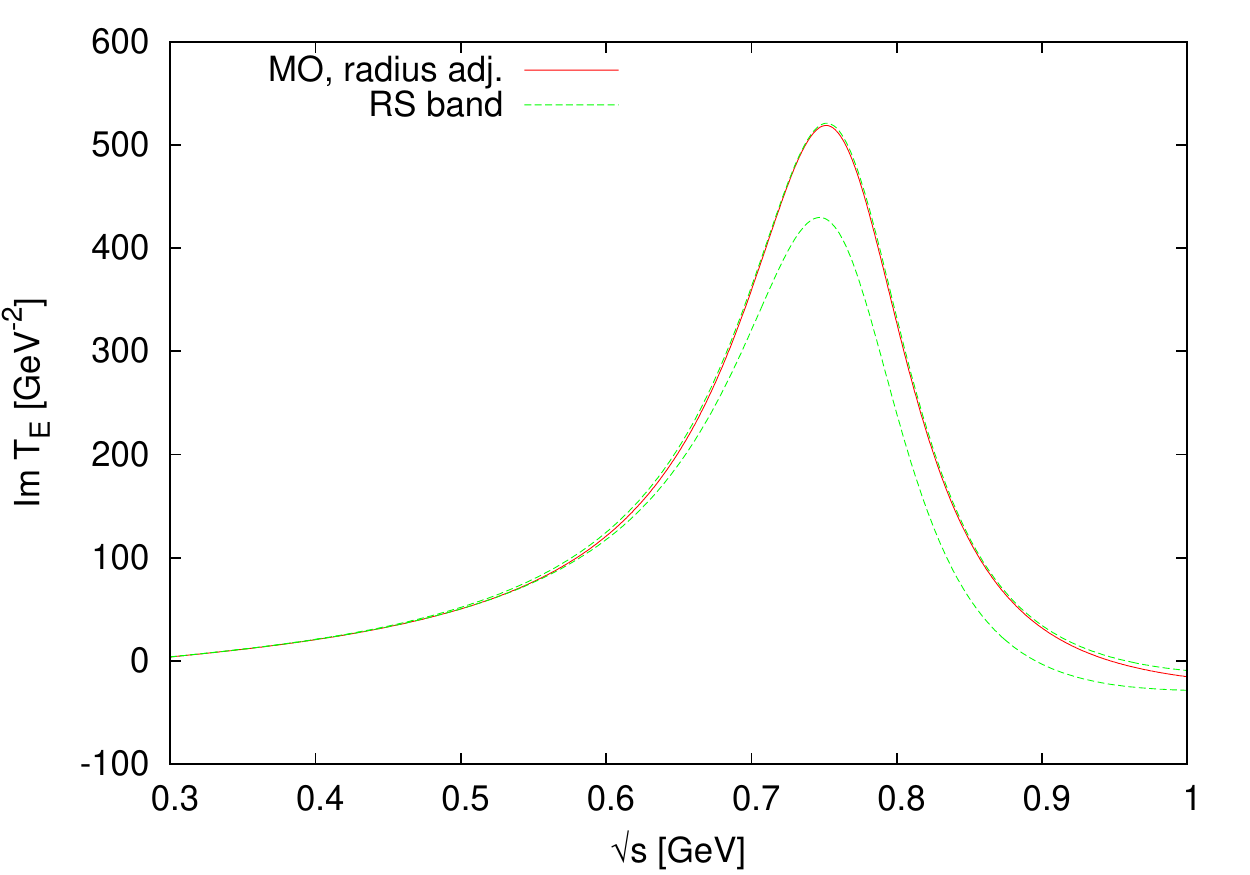}
  \end{minipage}   
  \caption{Real (top) and imaginary (bottom) part of the electric amplitude comparing RS to MO with $\Lambda = 1.8\,$GeV, 
    nucleon and Delta exchange and a constant that is adjusted to the (isovector) electric radius.}
  \label{fig:compReImTE}
\end{figure}

Cancelations of large terms enhance the 
sensitivity to small(er) corrections. Of course, LO=NLO nucleon+Delta $\chi$PT is an approximation. (This remark refers now to 
chiral corrections, not to large-$N_c$ corrections.) In fact convergence problems of nucleon $\chi$PT have also been observed 
in \cite{Becher:2001hv,Hoferichter:2015tha,Hoferichter:2015hva}. 
Thus it might be worth to explore an alternative approach already mentioned in 
section \ref{sec:chipt} (and essentially used also in the magnetic sector). 
Keeping Born and Delta exchange, but leaving $P_{0,E}$ as a free parameter, I can use the experimental value for the electric 
radius to determine $P_{0,E}$ from (\ref{eq:dispbasicradiifitP0}). What is needed in addition to the 
sum of (\ref{eq:polyelborn}), (\ref{eq:P0WT}), (\ref{eq:P0resE2}) is $\Delta P_{0,E} \approx 29.\,$GeV$^{-2}$. This value is 
smaller than each of the values of the separate contributions (\ref{eq:polyelborn}), (\ref{eq:P0WT}), and (\ref{eq:P0resE2}), 
but quite important for the total budget.
The results for the scattering amplitude $T^{\rm MO}_E$, given by (\ref{eq:tmandel}), 
can be compared to the results from the dispersive 
RS analysis of \cite{Hoferichter:2015hva}. This comparison is shown in figure \ref{fig:compReImTE}. 
Again an impressive agreement is observed given the simplicity of the input. It appears that the relevant physics 
contained in these helicity flip (magnetic) 
and non-flip (electric) p-wave subthreshold amplitudes is captured rather well by nucleon and Delta exchange 
unitarized by the MO method and accompanied by one subtraction constant per channel.

\begin{figure}[h!]
  \centering
  \begin{minipage}[c]{0.48\textwidth}  
    \includegraphics[keepaspectratio,width=\textwidth]{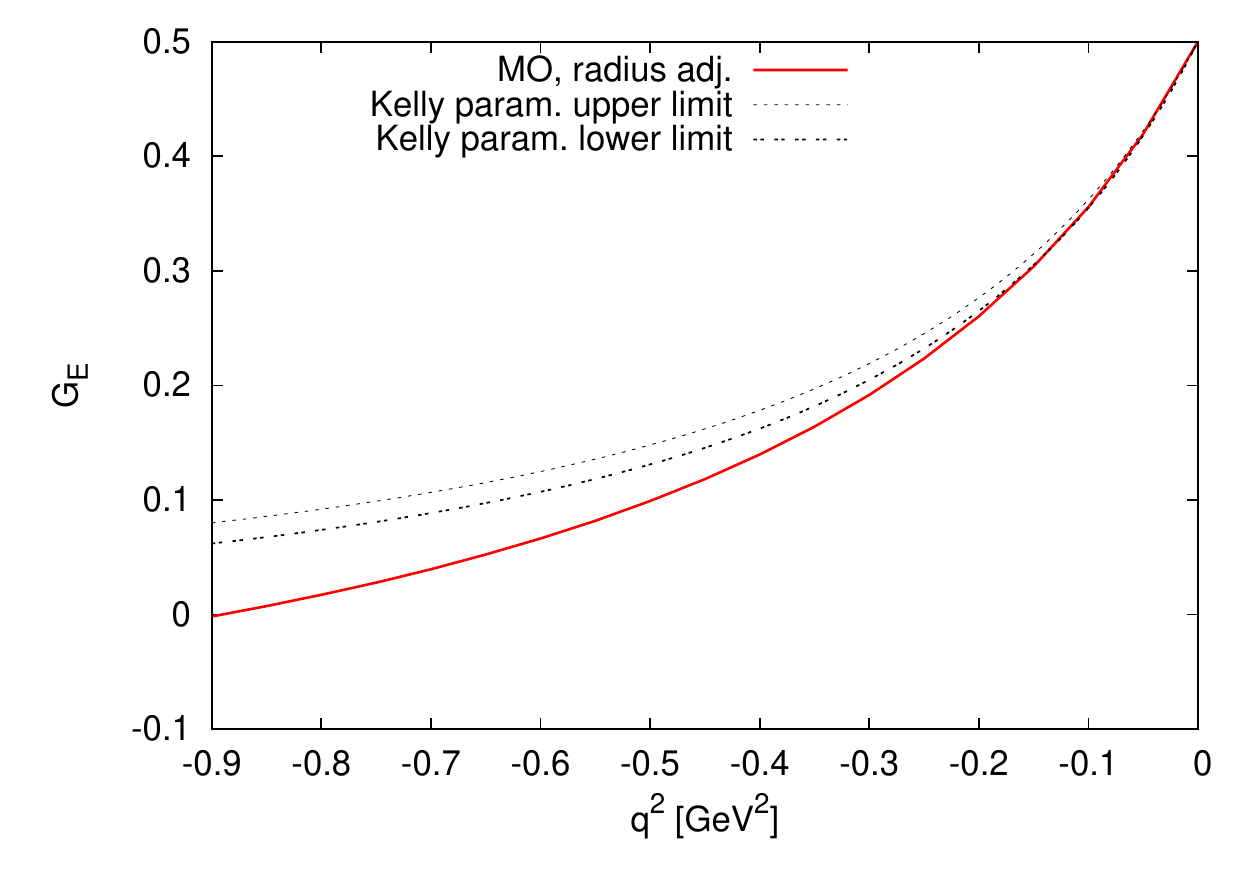}
  \end{minipage}   
  \caption{The (isovector part of the) electric form factor of the nucleon in the spacelike region from MO with nucleon and Delta
    exchange and two subtraction constants (for the MO and the form factor integral) that are adjusted to the isovector 
    charge and corresponding radius \cite{pdg}. 
    The Kelly parametrization
    \cite{Kelly:2004hm} is obtained from a fit to proton and neutron data.}
  \label{fig:GEspace}
\end{figure}

Finally the resulting (isovector) electric form factor in the spacelike region is depicted in figure \ref{fig:GEspace}.
A fair agreement is achieved, though not quite as satisfying as for the magnetic case. Obviously the tension to $\chi$PT requires
further investigations.

The results of the present work suggest that a low-energy isovector form factor of the transition 
from $A$ to $B$ can be calculated from a dispersion relation using the dominant two-pion inelasticity and the solution of the 
MO equation to account for pion rescattering. The required input are the dominant left-hand cut structures, 
which might be approximated by tree-level hadron exchange diagrams in the $s$ and $u$ channel of the reaction $\pi A \to \pi B$. 
This is similar in spirit to \cite{GarciaMartin:2010cw,Kang:2013jaa,Kubis:2015sga}.
As further experimental input the value of the form factor and its slope at the photon point are needed to pin down the 
subtraction constants.
A natural application of this framework are Dalitz decays $A \to B \, e^+ e^-$. Concrete examples are 
$(A,B)=(\Delta,N)$, $(\Sigma,\Lambda)$ \cite{Granados:2017cib}, $(\Sigma^*,\Lambda)$, $(\Sigma_c,\Lambda_c)$. These topics 
will be addressed in the future.
In addition it might be worth to explore if the deviations between the MO and N/D unitarization schemes are mitigated once 
the input is extended beyond NLO by calculating the required pion-nucleon scattering amplitudes at one-loop accuracy 
of relativistic nucleon+Delta $\chi$PT.

\begin{acknowledgement}
I thank Martin Hoferichter and Bastian Kubis for many valuable discussions and for providing the results of their 
dispersive Roy-Steiner equations. I also thank Emilie Passemar for inspiring discussions and creative suggestions  
how to make further use 
of the formalism developed here. 
\end{acknowledgement}

\appendix

\section{Comparing conventions for pion-nucleon amplitudes}
\label{sec:app0}

The reduced amplitudes $T_{M/E}$ used in \cite{Granados:2017cib} and here are related to the corresponding amplitudes of 
\cite{PhysRev.117.1603,Hoferichter:2016duk} via
\begin{eqnarray}
  \label{eq:matchbonn}
  f^1_-(s) = \frac{\sqrt{2}}{12\pi} \, T_M(s) \,, \qquad f^1_+(s) = \frac{m_N}{12\pi} \, T_E(s) \,.
\end{eqnarray}

For practitioners it might be helpful to compare also the expressions for the amplitudes before projecting on the p-wave. 
In \cite{Granados:2017cib} and here the formal reaction baryon plus antibaryon to two pions is studied.
In \cite{PhysRev.117.1603} it is the time reversed reaction 
while in \cite{Becher:2001hv,Hoferichter:2016duk} it is elastic pion-nucleon scattering. Correspondingly, the
independent variables are $s$ and scattering angle in \cite{Granados:2017cib} and $t$ and scattering angle 
in \cite{Becher:2001hv,Hoferichter:2016duk}. Also in \cite{PhysRev.117.1603} the variable $t$ is used. There it denotes the 
square of the invariant mass of the two-pion system. Thus it is $s$ in \cite{Granados:2017cib} and here what is called $t$ in 
\cite{PhysRev.117.1603,Becher:2001hv,Hoferichter:2016duk}.

In principle, one could imagine that a slightly larger complication 
than just rewriting $s$ to $t$ could emerge when turning from baryon-antibaryon spinor structures 
$\bar v \ldots u$ to $\bar u \ldots u$.
Depending on the definitions of $v$ and $u$ an analytic continuation might not lead from $v$ to $u$, but to $u$ 
times a sign or phase.
However, such problems are avoided in the formalism used in \cite{Granados:2017cib} and here. The reduced amplitudes are 
deduced from ratios where the convention ambiguities drop out. 

The expression that enters the projection formula for the 
magnetic (helicity-flip) sector is
\begin{eqnarray}
  \frac{{\cal M}(s,\theta,+1/2,-1/2)}{\bar v(-p_z,-1/2) \, \gamma^1 \, u(p_z,+1/2) \; p_{\rm c.m.}} 
  \label{eq:magB-}
\end{eqnarray}
the corresponding one for the electric (non-flip) sector is
\begin{eqnarray}
  \label{eq:elcompl}
  \frac{{\cal M}(s,\theta,+1/2,+1/2)}{\bar v(-p_z,+1/2) \, \gamma^3 \, u(p_z,+1/2) \; p_{\rm c.m.}}  \,.
\end{eqnarray}
For details see \cite{Granados:2017cib}. There the Feynman amplitude ${\cal M}$ is always decomposed into a structure 
that is proportional to 
$\bar v \, u$ and another one that is proportional to $\bar v \, \gamma^\mu k_\mu u$. For the following matching procedure it 
is helpful to recall that $k=p_+-p_-$ is the difference of 
pion momenta, chosen to lie in the $x$-$z$ plane. 

I start with the decomposition
\begin{eqnarray}
  \label{eq:decompAB}
  {\cal M} =: A_C \, \bar v \, u - \frac12 \, B_C \, \bar v \, \gamma^\mu k_\mu u  
\end{eqnarray}
where $A_C$ and $B_C$ can be easily read off from the expressions given in \cite{Granados:2017cib} and translated to the 
nucleon case via the replacement rules specified at the beginning of section \ref{sec:chipt}.
I will show now that the two scalar structures $A_C$ and $B_C$ defined in this way coincide with $A^-$ and $B^-$ 
from \cite{PhysRev.117.1603,Becher:2001hv,Hoferichter:2016duk}, respectively (except for calling $s$ then $t$). 
The index $C$ could be regarded as referring to crossing or to the first name of the 
first author of \cite{Granados:2017cib}. 

For the magnetic sector only $B_C$ contributes. To evaluate the ratio (\ref{eq:magB-}) one needs 
$k_1 = -2 \, p_{\rm c.m.} \sin\theta$. Then the formula for $T_M$, equation (23) in  \cite{Granados:2017cib}, takes the form
\begin{eqnarray}
  T_M(s) = \frac34 \, \int\limits_0^\pi d\theta \, \sin^3(\theta) \, B_C  \,.
  \label{eq:projMBC}  
\end{eqnarray}
Identifying $B_C$ with $B^-$ this fits exactly to the relation between $f^1_-$ and $B^-$ from 
\cite{PhysRev.117.1603,Hoferichter:2016duk}.

In the electric sector both $A_C$ and $B_C$ contribute. To evaluate (\ref{eq:elcompl}) it is helpful to use 
$k_3 = - 2 \, p_{\rm c.m.} \cos\theta$ and the relation
\begin{eqnarray}
  && \bar v(-p_z,+1/2) \, u(p_z,+1/2) \nonumber \\ 
  && = - \frac{p_z}{m_N} \, \bar v(-p_z,+1/2) \, \gamma^3 \, u(p_z,+1/2) \,.
  \label{eq:helunit}
\end{eqnarray}
Then the formula for $T_E$, equation (22) in  \cite{Granados:2017cib}, becomes
\begin{eqnarray}
  \label{eq:projEAB}  
  && T_E(s) \\ 
  && = \frac32 \, \int\limits_0^\pi d\theta \, \sin\theta \, 
  \left( -\frac{p_z}{m_N \, p_{\rm c.m.}} \, A_C + B_C \, \cos\theta \right) \, \cos\theta  \,.  \nonumber 
\end{eqnarray}
The relation between $f^1_+$ and $A^-$, $B^-$ from \cite{PhysRev.117.1603,Hoferichter:2016duk} is recovered 
if one identifies $A_C$ with $A^-$ (and again $B_C$ with $B^-$). 

As further cross-checks of my calculations I have explicitly compared my results for the Born terms with 
\cite{Hoferichter:2016duk} and for the Weinberg-Tomozawa term (\ref{eq:P0WT}) and the $c_4$ term (\ref{eq:P0NLO}) 
with \cite{Becher:2001hv} (details not shown here).

\section{Carrying a constant through the MO formalism}
\label{sec:app1}

In this appendix I compare left- and right-hand side of (\ref{eq:non-eq}) for the case that $K$ is just a 
constant; see also \cite{Kang:2013jaa}. 
The purpose is to demonstrate that it matters if $P_0$ is kept as a part of $K$ or is treated separately. 

The Omn\`es function (\ref{eq:omnesele}) scales like $1/s$ for large $s$. Essentially this is 
achieved by the phase shift $\delta$ approaching $\pi$ in the same limit. The inverse of the Omn\`es function grows linearly 
with $s$. Therefore a quadratic dispersion relation can be established \cite{Kang:2013jaa}:
\begin{equation}
  \label{eq:dispinvomnes}
  \Omega^{-1}(s) = 1 - \dot\Omega(0) \, s + s^2 \, \int\limits_{4m_\pi^2}^\infty \, \frac{ds'}{\pi} \, 
  \frac{{\rm Im}\Omega^{-1}(s')}{(s'-s-i \epsilon) \, {s'}^2}   \,.
\end{equation}

Instead of a completely general study I will focus on one particular high-energy behavior of $\Omega$, namely when 
the imaginary part of $\Omega^{-1}$ approaches a constant for large $s$. This happens if $\sin\delta(s) \sim 1/s$ for large $s$.
Then the dispersive integral in 
(\ref{eq:dispinvomnes}) can be rewritten:
\begin{eqnarray}
  \label{eq:dispinvomnes2}
  && s^2 \, \int\limits_{4m_\pi^2}^\infty \, \frac{ds'}{\pi} \, 
  \frac{{\rm Im}\Omega^{-1}(s')}{(s'-s-i \epsilon) \, {s'}^2}
  = \nonumber \\
  && - s \int\limits_{4m_\pi^2}^\infty \, \frac{ds'}{\pi} \, 
  \frac{{\rm Im}\Omega^{-1}(s')}{{s'}^2}
  + s \int\limits_{4m_\pi^2}^\infty \, \frac{ds'}{\pi} \, 
  \frac{{\rm Im}\Omega^{-1}(s')}{(s'-s-i \epsilon) \, {s'}}   \,.  \nonumber \\  &&
\end{eqnarray}
Thus one finds
\begin{eqnarray}
  \label{eq:dispinvomnes3}
  s \int\limits_{4m_\pi^2}^\infty \, \frac{ds'}{\pi} \, 
  \frac{{\rm Im}\Omega^{-1}(s')}{(s'-s-i \epsilon) \, {s'}} 
  = \Omega^{-1}(s) - 1 + L \, s  
\end{eqnarray}
with the constant 
\begin{eqnarray}
  \label{eq:dispinvomnes4}
  L := \dot\Omega(0) + \int\limits_{4m_\pi^2}^\infty \, \frac{ds'}{\pi} \, 
  \frac{{\rm Im}\Omega^{-1}(s')}{{s'}^2}  \,.
\end{eqnarray}
For a constant $K$ the last expression in (\ref{eq:non-eq}) turns to $K-K \, L \, s$. In other words it makes a difference 
if a constant term in the low-energy expression is just identified with $P_0$ or carried through the MO machinery with a once 
subtracted dispersion relation. 

The difference is of order $s$ which is beyond the NLO approximation that is used throughout
this work. Note that in contrast to pionic $\chi$PT, where the respective next order in the power expansion is suppressed 
by $s$, baryonic $\chi$PT order by order receives relative corrections of order $\sqrt{s}$ \cite{Scherer:2012xha}. 

\bibliography{lit}{}
\bibliographystyle{epj}
\end{document}